\documentclass[preprint,preprintnumbers,floatfix,
showkeys,superscriptaddress]{revtex4}
\usepackage{amsmath,amssymb,amsfonts}
\usepackage{graphicx}
\usepackage{color}
\usepackage{hyperref}

\allowdisplaybreaks

\newcommand{\be}{\begin{equation}}
\newcommand{\ee}{\end{equation}}
\newcommand{\bea}{\begin{eqnarray}}
\newcommand{\eea}{\end{eqnarray}}

\begin{document}



\title{Limiting symmetry energy elements from empirical evidence}

\author{B. K. Agrawal}
\email{bijay.agrawal@saha.ac.in}
\address{Theory Division, Saha Institute of Nuclear Physics, 1/AF 
 Bidhannagar, Kolkata 700064, India.}  
\address{Homi Bhabha National Institute, Anushakti Nagar, Mumbai 400094, India.}
\author{S.K. Samaddar}
\email{santosh.samaddar@saha.ac.in}
\address{Theory Division, Saha Institute of Nuclear Physics, 1/AF 
 Bidhannagar, Kolkata 700064, India.}  
\author{J.N. De}
\email{jn.de@saha.ac.in}
\address{Theory Division, Saha Institute of Nuclear Physics, 1/AF 
 Bidhannagar, Kolkata 700064, India.}  
\author{C. Mondal}
\email{chiranjib.mondal@saha.ac.in}
\address{Theory Division, Saha Institute of Nuclear Physics, 1/AF 
 Bidhannagar, Kolkata 700064, India.}  
\address{Homi Bhabha National Institute, Anushakti Nagar, Mumbai 400094, India.}
\author{Subhranil De}
\email{subde@ius.edu}
\address{Department of Physical Sciences, Indiana University SouthEast,
IN 47150, USA}

\begin{abstract} 
In the framework of an equation of state (EoS) constructed from a momentum
and density-dependent finite-range two-body effective interaction,
the quantitative magnitudes of the different symmetry elements
of infinite nuclear matter are explored. The parameters
of this interaction  are determined from well-accepted characteristic
constants associated with homogeneous nuclear matter. 
The symmetry energy coefficient $a_2$, its density slope $L_0$, the
symmetry incompressibility $K_\delta $ as well as the density dependent
incompressibility $K(\rho )$ evaluated with this EoS are seen to be 
in good harmony with those obtained from other diverse perspectives.
The higher order symmetry energy coefficients $a_4,~a_6$ etc are seen to be
not very significant in the domain of densities relevant to finite nuclei,
but gradually build up at supra-normal densities. The analysis carried with a Skyrme-inspired energy
density functional obtained with the same input values for the empirical
bulk data associated with nuclear matter yields nearly the same results.

\end{abstract}



\maketitle

\section{Introduction}

Much attention has recently been drawn to a precise understanding of
the different aspects of nuclear symmetry energy. For nuclei with extreme
isospins they are the predominant factors in determining their stability
and the nucleon distributions therein \cite{Myers69,Myers80,Agrawal12}.
In astrophysics, they have seminal influence on the size, critical
composition and maximum mass of neutron stars \cite{roberts12,steiner08}.
The dynamical evolution of the core collapse of a massive star and the
associated explosive nucleosynthesis \cite{steiner05,janka07} also
depend sensitively on them.

Nuclear symmetry energy is the energy cost in converting asymmetric nuclear
matter to a symmetric one. It is defined as

\bea
\label{esym1}
e_{sym}(\rho ,\delta )=e(\rho ,\delta )-e(\rho ,\delta =0),
\eea
where $e$ is the energy per nucleon of nuclear matter,
$\delta =(\rho_n - \rho_p )/(\rho_n +\rho_p )$ is the nuclear asymmetry
and $\rho_n$ and $\rho_p$ are the neutron and proton densities with
$\rho_n +\rho_p = \rho $. Expanding $e(\rho ,\delta)$ in powers 
of $\delta $ around $\delta =0$ and keeping only the even powers of
$\delta$ (because of charge symmetry), one has
\bea
\label{esym2}
e_{sym}(\rho ,\delta )=a_2\delta^2+a_4\delta^4+a_6\delta^6 +\cdots ,
\eea
where
\bea
\label{a2}
a_2=\biggl [\frac {1}{2} \frac{\partial^2e_{sym}(\rho ,\delta )}
{\partial \delta^2} \biggr ]_{\delta =0} ,
\eea

\bea
\label{a4}
a_4=\biggl [\frac {1}{4!} \frac{\partial^4e_{sym}(\rho ,\delta )}
{\partial \delta^4} \biggr ]_{\delta =0} ,
\eea
\bea
\label{a6}
a_6=\biggl [\frac {1}{6!} \frac{\partial^6e_{sym}(\rho ,\delta )}
{\partial \delta^6} \biggr ]_{\delta =0} ,
\eea
and so on.

Traditionally since the days of Bethe and Weiz\" acker \cite{Weizacker35,
Bethe36}, only the first term in the expansion (\ref{esym2})
has been considered for symmetry energy. 
If so, the coefficient of symmetry energy as  obtained 
from the double derivative of $e_{sym}(\rho ,\delta )$ 
is true for any value of $\delta$ and the
symmetry energy can then be taken as 
\bea
\label{esym3}
e_{sym}(\rho )=e(\rho ,\delta =1)-e(\rho ,\delta =0),
\eea
which has been resorted to by some in its definition \cite{
Natowitz10}. At low density when matter becomes clusterized, the
two definitions given by Eqs. (\ref{esym1}) and (\ref{esym3}) show
different behavior \cite{De10}. For homogeneous nuclear matter,
however, up to around 
$\rho_0$, the symmetry energy $e(\rho ,\delta )$
shows nearly a perfect linearity in $\delta^2$ \cite{De10,Chen09}
in microscopic calculations with different energy density functionals
(EDF) used to explain nuclear properties corroborating the 
Bethe-Weiz\" acker ansatz.

Even if terms beyond $\delta^2$ in Eq.(\ref{esym2}) are unimportant for
 accounting the symmetry energy at normal density, at supra-normal
densities, they can not be ignored as has recently been shown in 
calculations with Skyrme EDFs \cite{Constantinou14}. Mean-field
calculations in a nonlinear relativistic framework \cite{Cai12}
suggest also such an outcome. These higher order terms are important
to reasonably describe the proton fraction of $\beta$-stable nuclear
matter at high densities and the core-crust transition
density in neutron stars \cite{Seif14}. 

In contrast to the generally accepted idea that terms beyond
$\delta^2$ are relatively unimportant in symmetry energy at normal density,
a recent analysis of the double differences of 'experimental' symmetry
energies of neighboring nuclei \cite{Jiang14, Jiang15} indicates that the
higher order terms in symmetry energy for finite nuclei  may be sizeable even at saturation
density. However, no firm conclusions could be drawn because of 
the model  dependence
in evaluating the nuclear masses. 
With the standard Skyrme energy density functionals the fourth order term (with $\delta^4$) 
comes out to be negative from the binding energy formula \cite{Wang15}, whereas the latest 
Weiz\"acker-Skyrme formula and the extracted value \cite{Jiang14} from the experimental data 
suggest positive values for this coefficient.
In this context,  a
reexamination of the importance of the higher order terms in symmetry
energy for infinite nuclear matter  is  called for. The present 
communication is aimed towards that purpose.

Employing variants of the Bethe-Weiz\" acker mass formula, attempts were
made to extract the value of the symmetry energy 
of nuclear matter from the 
known experimental nuclear masses \cite{Royer08,Moller12}. The
symmetry energy of a finite nucleus has two components, the volume 
and the surface one. The volume term relates  to the symmetry
energy coefficient of infinite nuclear matter
at the saturation density $\rho_0$, the surface term comes from 
finite-size effects. Extraction
of the volume part of the nuclear symmetry energy 
from nuclear masses suffers some ambiguity because of the 
interference of the surface term. The nuclear binding energies may
be well represented, but the volume and surface symmetry terms
may vary over a considerable range \cite{Jiang14, Antonov16}, a large volume term
is compensated by a large surface term and vice versa. 

 Microscopic theories built out of effective two-nucleon interactions
\cite{Brack85} structured to explain selective experimental data have
not yet been able to completely address the problem of properly 
delineating the symmetry elements of nuclear matter from finite
nuclear properties. For example, both the relativistic NL3 interaction
\cite{Gambhir90} and the non-relativistic BSk24 \cite{Goriely13} give
very good fit to the nuclear masses, but the symmetry element 
$a_2$ at $\rho_0$  in the
former case is 37.4 MeV, in the latter case it is 30 MeV. The density
slope of symmetry energy at  $\rho_0$, namely
$L(\rho_0)$ ($=L_0$ defined as $3\rho_0 \frac{\partial a_2(\rho)}{\partial \rho}|_{\rho_0}$) 
varies even more significantly, $L_0$ = 46.4 MeV
for the BSk force, but is 118.5 MeV for the NL3 interaction. There
is thus no clear consensus on the values of the different symmetry
elements pertaining to nuclear matter from microscopic theories
\cite{Chen09}, though they are largely successful in fitting diverse
experimental data. 

Through the maze of different experimental facts and their theoretical analyses, some empirical
constants related to nuclear matter, however, have emerged that
seem to lie in nearly tight limits. They are 
the saturation density $\rho_0$ of symmetric nuclear matter
(SNM) and its energy per nucleon $e_0$ at that density 
\cite{Chen09,Dutra12,Moller12,Akmal97,Baldo13,Myers98}. 
The nuclear incompressibility $K_0$ of SNM at $\rho_0$
have been progressively refined  and is now relatively
well constrained \cite{Shlomo06,Khan12,De15}. We choose these
empirical data as benchmarks to fix the isoscalar part of the 
effective interaction that would be used to explore nuclear 
matter properties. For the proper feel of the isovector component,
we exploit an empirically observed characteristic of pure neutron
matter (PNM). From a large number of 'best-fit' 
EDFs \cite{Brown13} built in the Skyrme framework, it has been seen 
that the value of energy per particle for PNM at density $\rho $
=0.1 fm$^{-3}$ is practically the same, $e_n \sim $10.9 MeV.
This is another benchmark we take recourse to. Incidentally,
this value of $e_n$ is in extremely good consonance with that obtained for PNM
from the most realistic microscopic potential model calculations of
Akmal and Pandharipande \cite{Akmal97} and Akmal, Pandharipande
and Ravenhall \cite{Akmal98}. The agreement of this value for $e_n$ is also excellent
with that obtained from the ab initio advanced microscopic calculation by 
Baldo {\it et al} \cite{Baldo13} within the Kohn-Sham density functional framework. 
The neutron-matter data is chosen so that extrapolation
to highly asymmetric matter becomes reliable. 
In addition to the above benchmark empirical data, 
the value of the effective
mass of the nucleon $m^*(\rho_0)/m$ for SNM at saturation 
density is taken as a given input. The
parameters describing the effective interaction can then be
calculated from the given conditions. The value of $m_0^*/m$
(from now on, we write $m_0^*$ for $m^*(\rho_0)$)
is constrained such that the observed 
maximum mass $M_{max}^{NS}$ of the neutron star
\cite{Demorest10,Antoniadis13} is in consonance with the calculated result.

To build the EDF, we confine ourselves in the non-relativistic
framework. We start with a density and momentum dependent finite-range
effective two-body interaction in the modified Seyler-Blanchard
(SBM) prescription. This simple interaction with few parameters
has been applied earlier to evaluate successfully many a nuclear
properties \cite{Blocki77,De96}.  A variant of this interaction has
also been used by Myers and Swiatecki \cite{Myers90} to calculate
nuclear masses, nuclear deformations, charge distributions etc. and
is seen to reproduce these properties very well.  Calculations of EDF
with empirical nuclear constants as base have been attempted earlier
\cite{Bandyopadhyay90,Alam14}. In Ref. \cite{Bandyopadhyay90}, the
SBM prescription for the form of the effective interaction was taken,
in Ref. \cite{Alam14}, the interaction was of the zero-range Skyrme
class. The present calculations have been done in the same spirit,
however, the chief difference with the earlier ones is that previously
the parameters of the interactions were calculated with the symmetry
energy element  $a_2(\rho_0)$ being kept fixed at a predetermined value
and that it was further equated with $e_{sym}(\rho_0)$.  This masked the
higher order effects in the asymmetry parameter $\delta $.  Moreover,
in the cases so mentioned, attempts were not made to find the maximum mass
of the neutron star in relation to the interaction parameters.

  The value of $e_0$, the energy per nucleon for SNM at 
$\rho_0$ is taken as $e_0 =-$16.0 $\pm $0.2 MeV with $\rho_0
=$ 0.155 $\pm $0.008 fm $^{-3}$. 
There is a still no clear consensus on the strict bounds on $e_0$ or $\rho_0$. 
For example, some models lead to somewhat lower values for $e_0$ \cite{Dieperink09, Wang10, Pomorski03}, 
we adhere to the value obtained from the recent version of the finite range droplet model (FRDM) 
\cite{Moller12} that agrees better with the new mass database. 
The incompressibility $K_0$ is
taken to be 215 $\pm $25 MeV \cite{De15}.  This is somewhat lower than the value of $K_0 = 230\pm
40$ MeV as inferred in Ref. \cite{Khan12}, but is consistent with the incompressibility of
symmetric nuclear matter and its density slope at the sub-saturation crossing density $\rho_c$ as
explained in Ref. \cite{De15}. 
The value of the per particle energy of PNM at
density $\rho =$0.1 fm $^{-3}$ is taken as $e_n=$
10.9 $\pm $0.5 MeV \cite{Brown13}.

  The paper is organized as follows. In Sec. II, we review the elements
of theory. In Sec. III, the results and discussions are presented.
The concluding remarks are drawn in Sec. IV.

\section{Theoretical details}

In the following, we describe the form of the effective two-body interaction
and briefly outline the procedure for determining the parameters of this
interaction from given empirical nuclear data. From the EDF constructed
with this interaction, the different isovector elements 
pertaining to nuclear matter are then calculated, the question of the lower
limit of the maximum mass of the neutron star
$M_{max}^{NS}$ is further addressed. 

\subsection{Effective interaction and the nuclear EoS}

The Seyler-Blanchard effective interaction \cite{Seyler61} in the
modified version \cite{Bandyopadhyay88} is taken to be of the form
\bea
\label{veff}
v_{eff}(r,p,\rho )=C_{l,u}\bigl [v_1(r,p)+v_2(r,\rho ) \bigr ],
\eea

\bea
\label{v1}
v_1= -(1-\frac{p^2}{b^2})f({\bold r_1},{\bold r_2}),
\eea

\bea
\label{v2}
v_2 = d^2\bigl [ \rho (r_1)+\rho(r_2)\bigr ]^\alpha f({\bold r_1},
{\bold r_2}),
\eea
with
\bea
\label{fr}
f({\bold r_1},{\bold r_2})=\frac{e^{-|{\bold r_1}-{\bold r_2}|/a}}
{|{\bold r_1}-{\bold r_2}|/a} .
\eea
 Here the subscripts $l$ and $u$ to the interaction strength $C$ refer
to like pair ($nn$, or $pp$) or unlike pair ($np$) interaction,
$a$ is its spatial range and $b$ the strength of repulsion in its 
momentum dependence. The relative coordinate is $r=|{\bold r_1}-
{\bold r_2}|$, the relative momentum is $p=|{\bold p_1}-{\bold p_2}|$,
with 1 and 2 referring to the two interacting nucleons, 
$\rho (r_1 )$ and
$\rho (r_2)$ being the densities at their sites. The parameters $d$ and
$\alpha$ are the measures of the strength of the density dependence
in the interaction.

To construct the EoS from the effective interaction, one needs to know
the occupation probability $n_\tau ( p,T)$ where $T$ is the
temperature and $\tau $ referring to the isospin index (neutrons 
or protons). The
self-consistent occupation probability in asymmetric nuclear matter
at $T$ is obtained by minimizing the thermodynamic potential $G$
\bea
\label{gp}
G=E-TS-\sum_\tau \mu_\tau N_\tau ,
\eea
where $E$ and $S$ are the total energy and entropy of the system,
and $\mu_\tau$ and $N_\tau$ are the respective chemical potentials
and total numbers of the isospin species.
Following ref. \cite{Bandyopadhyay90}, the
minimization of the thermodynamic potential 
with this interaction leads to the expression for the
occupation probability as

\bea 
\label{occ1}
n_\tau (p,T)=\biggl [1+e^{\bigl\{(\frac{p^2}{2m_\tau^*}+V_\tau^0+
V_\tau^2-\mu_\tau)/T\bigr\}}\biggr ]^{-1}.
\eea
Here $m_\tau^*$ is the nucleon effective mass.
The momentum-dependent
part of the single-particle potential $V_\tau^0 + p^2V_\tau^1$ 
defines the effective mass  as
\bea
\label{mstar}
m_\tau^*=\bigl [\frac{1}{m_\tau} + 2V_\tau^1 \bigr]^{-1},
\eea
where $m_\tau $ is the bare nucleon mass. The quantity $V_\tau^2$ is
the rearrangement energy that vanishes for density-independent effective
interactions.

Recently, symmetry energy and associated properties of finite nuclei have been
studied at finite temperature \cite{Antonov17}. 
In this paper we are dealing with the properties of nuclear matter
in the ground state ($T=$0).
In the limit T$\rightarrow $ 0, the occupation function $n_\tau ({\bold p})$
becomes the Heaviside theta function,
\bea
\label{occ2}
n_\tau (p)=\Theta [P_{F,\tau }- p ] ,
\eea
where  the Fermi momentum $P_{F,\tau}$ given by
\bea
\label{pfm}
\frac{P_{F,\tau}^2}{2m_\tau^*}= \mu_\tau -V_\tau^0 -V_\tau^2 ,
\eea
is related to density as $P_{F,\tau }=(3\pi^2\rho_\tau)^{1/3}\hbar $.
The expressions for different parts of the single-particle potential and
the rearrangement term, at zero temperature are 
given as \cite{Bandyopadhyay90}
\bea
\label{vt0}
V_\tau^0&=&-4\pi a^3\{1-d^2(2\rho )^\alpha\} (C_l\rho_\tau +C_u\rho_{-\tau})
  \nonumber \\
&&+\frac{4}{5\pi}(3\pi^2)^{5/3}\frac{a^3\hbar^2}{b^2} (C_l\rho_\tau^{5/3}
+C_u\rho_{-\tau}^{5/3}) ,
\eea
\bea
\label{vt1}
V_\tau^1 = \frac{4\pi a^3}{b^2}(C_l\rho_\tau+C_u\rho_{-\tau }) ,
\eea
\bea
\label{vt2}
V_\tau^2&=& 4\pi a^3d^2(2\rho )^{\alpha -1}\alpha 
[(C_l\rho_\tau+C_u\rho_{-\tau})\rho_\tau \nonumber \\
&&+(C_l\rho_{-\tau}+C_u\rho_\tau )\rho_{-\tau }].
\eea
In Eqs. (\ref{vt0}) -  (\ref{vt2}), if $\tau$ refers to proton, $-\tau$ 
refers to neutron and vice versa.
The density is given by
\bea
\label{rhot1}
\rho_\tau =\frac{2}{h^3}\int_0^{P_F} n_{\tau }(p)d{\bold p} \nonumber \\
=\frac {2\sqrt { 2}}{3\pi^2\hbar^3}(m_{\tau }^*)^{3/2}(\mu_\tau -V_\tau^0 -
V_\tau^2)^{3/2}.  
\eea
 The total energy of nuclear matter per nucleon is then written as
\cite{De15}
\bea
\label{etot1}
e(\rho,\delta)=\frac{1}{\rho} \sum_{\tau} \rho_\tau \bigl [ 
\frac{3P_{F,\tau}^2}{20m_\tau} \bigl (1+\frac{m_\tau}{m_\tau^*} \bigr )
+\frac{1}{2}V_\tau^0 \bigr ] ,
\eea
and the total pressure is
\bea
\label{pres1}
P=\sum_\tau \rho_\tau \bigl [\frac{P_{F,\tau}^2}{2m_\tau}
\bigl (\frac{7}{10}\frac{m_\tau}{m_\tau^*}-\frac{3}{10} \bigr )
+\frac{1}{2}V_\tau^0 +V_\tau^2 \bigr ].
\eea
The expressions (\ref{etot1}) and (\ref{pres1}), for SNM reduce to 
\bea
\label{etot2}
e(\rho ,\delta =0)=\frac{3}{10}\bigl [\frac{P_F^2}{2m} 
(1+\frac{m}{m^*(\rho )}) \bigr ]
+\frac{1}{2}V_0,
\eea
\bea
\label{pres2}
P(\rho ,\delta =0)= \bigl [\frac{P_F^2}{2m} \bigl (\frac{7m}
{10m^*(\rho )} -\frac{3}{10} \bigr ) +\frac{1}{2}V_0 +V_2 \bigr ]\rho,
\eea
where $P_F=(\frac{3\pi^2}{2})^{1/3}\hbar \rho^{1/3}$ is the Fermi momentum,
$V_0, V_1, V_2 $ are the single-particle potentials and $m^*(\rho )$ the
effective mass, all for SNM. In our calculations, we have taken
the bare neutron and proton masses to be equal ($m_\tau =m$).

\subsection{Determination of the interaction parameters and 
symmetry elements}

The effective interaction as given by 
Eqs.~(\ref{veff})-(\ref{fr}) contains six unknown 
parameters, $C_l, C_u, a, b, d $, and $\alpha $. Out of these, as we find
later, for infinite nuclear matter, the parameters $C_l, C_u$ and $a$ 
appear in combination as $C_la^3$ and $C_ua^3$. 
It is then effectively five unknown
parameters we need to determine. The given empirical data are the energy
per particle $e_0$ at the saturation density $\rho_0$ for SNM when pressure is zero,
its incompressibility coefficient $K_0$, and $e_n$, the energy per particle 
of neutron matter at $\rho =$0.1 fm$^{-3}$. In addition, we take the value of 
$m_0^*/m$ for SNM as a free input such that a close contact of the
calculated value of $M_{max}^{NS}$ from the EDF can be established with
the current observed value of $M_{max}^{NS}$ =2.01 $\pm $0.04 $M_{\odot}$.   
The quantities $e_0$ and $P(\rho_0)(=0)$ are obtained from Eqs.~(\ref{etot2})
and (\ref{pres2}) by setting $\rho =\rho_0 $. The incompressibility is
obtained from Eq.~(\ref{pres2}) as 
\bea
\label{comp1}
K_0=9\frac{dP}{d\rho }|_{\rho =\rho_0 },
\eea
which, after some algebraic manipulation, reduces to 
\bea
\label{comp2}
K_0&=&-3V_0+(9\alpha +3)V_2 +V_1 \bigl [10.8P_{F,0}^2 +4.5b^2 \nonumber \\
&&\times \{ (\alpha +1)d^2(2\rho_0)^{\alpha }-1 \}  \bigr ].
\eea
In Eq.~(\ref{comp2}), $P_{F,0}$ is the Fermi momentum at $\rho_0$.
The neutron matter energy at density $\rho_n$ can be obtained from Eq.
(\ref{etot1}) setting $\delta =1$ as
\bea
\label{en1}
e_n&=&\frac{3}{10m} (3\pi^2)^{2/3} \hbar^2\rho_n^{2/3} +C_la^3 \bigl [
\frac{12\pi }{5}(3\pi^2)^{2/3}\frac{\hbar^2}{b^2}\rho_n^{5/3} \nonumber \\
&&-2\pi \rho_n\{1-d^2(2\rho_n)^\alpha \} \bigr ].
\eea

\begin{table}
\caption{ \label{tab1}The parameters of the effective interaction (in MeV fm units)}
\begin{ruledtabular}
\begin{tabular}{cccccc}
$m_0^*/m$&  $C_la^3$& $C_ua^3$& $b$& $d$& $\alpha $\\
\hline
0.65& 471.9 & 1269.3&2430.6 & 0.982& 0.0193\\
\hline
0.75& 103.2 & 295.0 & 1477.4 & 0.942 & 0.1235\\
\end{tabular}
\end{ruledtabular}
\end{table}

From the four given empirical data and a chosen value of $m_0^*/m$,
the five unknown parameters of the interaction $C_la^3, C_ua^3,b,d$
and $\alpha $ can be determined (see Appendix A). Since we are
interested in properties of homogeneous nuclear matter, we do not
need to determine $C_l, C_u$ and $a$ separately. That can be done
if we take into consideration semi-infinite matter and put another
constraint, say, a given value of its surface energy.  The values of
the interaction parameters are given in Tab. \ref{tab1} for two values
of $m_0^*/m$, namely, 0.65 and 0.75. This choice of the effective
mass is consistent with the empirical values obtained from many recent
optical-model analyses \cite{Jaminon89,Li15}. Covariance analysis of
symmetry observables from heavy ion flow data \cite{Zhang15b,Coupland16}
would put the value of $m_0^*/m$ at $\sim$ 0.7-0.8, the situation
is, however, not unambiguous.

From Eqs. (\ref{esym2}), (\ref{vt0}) and (\ref{etot1}), 
the symmetry coefficients at a density $\rho $,
in terms of the potential parameters read as,
\bea
\label{a22}
a_2=\frac{P_F^2}{6m}+\frac{4\pi a^3\rho P_F^2}{3b^2}(2C_l-C_u) \nonumber \\
-\pi a^3 \rho \{1-d^2(2\rho)^{\alpha } \}(C_l-C_u),
\eea
\bea
\label{a42}
a_4=\frac{P_F^2}{162m}+\frac{4\pi a^3\rho P_F^2}{81b^2} (2C_u-C_l),
\eea
\bea
\label{a62}
a_6=\frac{7P_F^2}{4474m}+\frac{28 \pi a^3P_F^2}{10935b^2} (7C_u-2C_l).
\eea
The total density slope of symmetry energy $L_t(\rho )= 3\rho 
(\partial e_{sym}/\partial \rho )$ is obtained using Eq.~(\ref{esym3}) as,
\bea 
\label{ll1}
L_t(\rho )&=&2^{2/3}P_F^2(\frac{3}{5m}+\frac{12 \pi }{b^2} C_la^3\rho ) \nonumber \\ 
&&-6\pi C_la^3\rho \{ 1-d^2(2\rho )^{\alpha }(1+\alpha ) \}.
\eea
In the literature, the symmetry slope $L(\rho)$ has, however, been usually taken as 
\bea
L(\rho )=3\rho \frac{\partial a_2(\rho)}{\partial \rho},
\eea
which from Eq. (\ref{a22}) is evaluated as
\bea
L(\rho)&=&\frac{P_F^2}{3m}+\frac{20\pi a^3}{3 b^2} (2C_l-C_u)\rho P_F^2\nonumber \\
&&-3\pi a^3(C_l-C_u)\rho\left[\{1-d^2(2\rho)^{\alpha}\}\right. \nonumber\\
&&-\left.\alpha d^2(2\rho)^{\alpha}\right]
\eea

\section{Results and discussions}

From the wealth of diverse theoretical enterprises like the liquid drop
type models \cite{Myers96,Myers98,Moller12}, the microscopic ab-initio
or variational calculations \cite{Baldo13,Akmal97} or different Skyrme or
Relativistic mean field models (RMF) - all initiated to explain different
experimental data, we choose saturation density as
 $\rho_0$=0.155$\pm $0.008 fm$^{-3}$ and the energy per nucleon
for SNM  as $e_0 = -16.0 \pm 0.2$ MeV, respectively. The value of the
nuclear incompressibility $K_0$, obtained from the microscopic analysis of
isoscalar giant monopole resonances (ISGMR) in nuclei has gone through
several revisions \cite{Rutel05,Avogadro13,Niksic08} from its early
value of $K_0 \simeq 210\pm 30 $ MeV \cite{Blaizot80, Farine97}. Now,
with the understanding that the ISGMR centroid energy reflects better
the density dependence of the incompressibility \cite{Khan12,Khan10},
its value has been reassessed \cite{De15} to $K_0 \simeq 215 \pm 25$
MeV. For $K_0$, we choose this input value.  This is not much different
from the early value quoted.

 For the effective mass, as explained in Appendix B, the
minimum value with the given central values of the   empirical inputs
for this effective interaction is $(m_0^*/m)_{min} \sim 0.64$. We
keep $(m_0^*/m)$ as a free parameter above this value. We find,
as shown later, that a low value of $m_0^*/m$ explains better the
lower limit of $M_{max}^{NS}$, it increases with decreasing effective
mass. We therefore  fix the central value of  $m_0^*/m$ at 0.65,
close to the lower limit, with an uncertainty of $\pm 0.1$.  As already
mentioned, this value of the isoscalar effective mass is coincident
with that obtained recently \cite{Li15} from a global analysis of
nucleon-nucleus scattering data within an isospin-dependent optical
model. In finite nuclei, the effective mass is typically closer to unity
\cite{Jeukenne76,Brown13} because of its enhancement due to the coupling
of the single-particle motion to the surface vibrations, but this has
not been included in the optical model analysis \cite{Li15}.
  The value of the energy per particle for
neutron matter is taken to be $e_n=10.9 \pm 0.5$ MeV at $\rho $= 0.1
fm$^{-3}$ \cite{Brown13}. Out of several hundred Skyrme EDFs, sixteen
of them nicely reproduced a selected set of experimental nuclear matter
properties. They gave $e_n=11.4 \pm 1.0 $ MeV at $\rho $=0.1 fm$^{-3}$
for PNM, among them six 'best-fit' results gave a more restricted range
(10.9$\pm$0.5 MeV) which we have chosen for $e_n$.

\begin{figure}
\resizebox{6.0in}{!}{\includegraphics[]{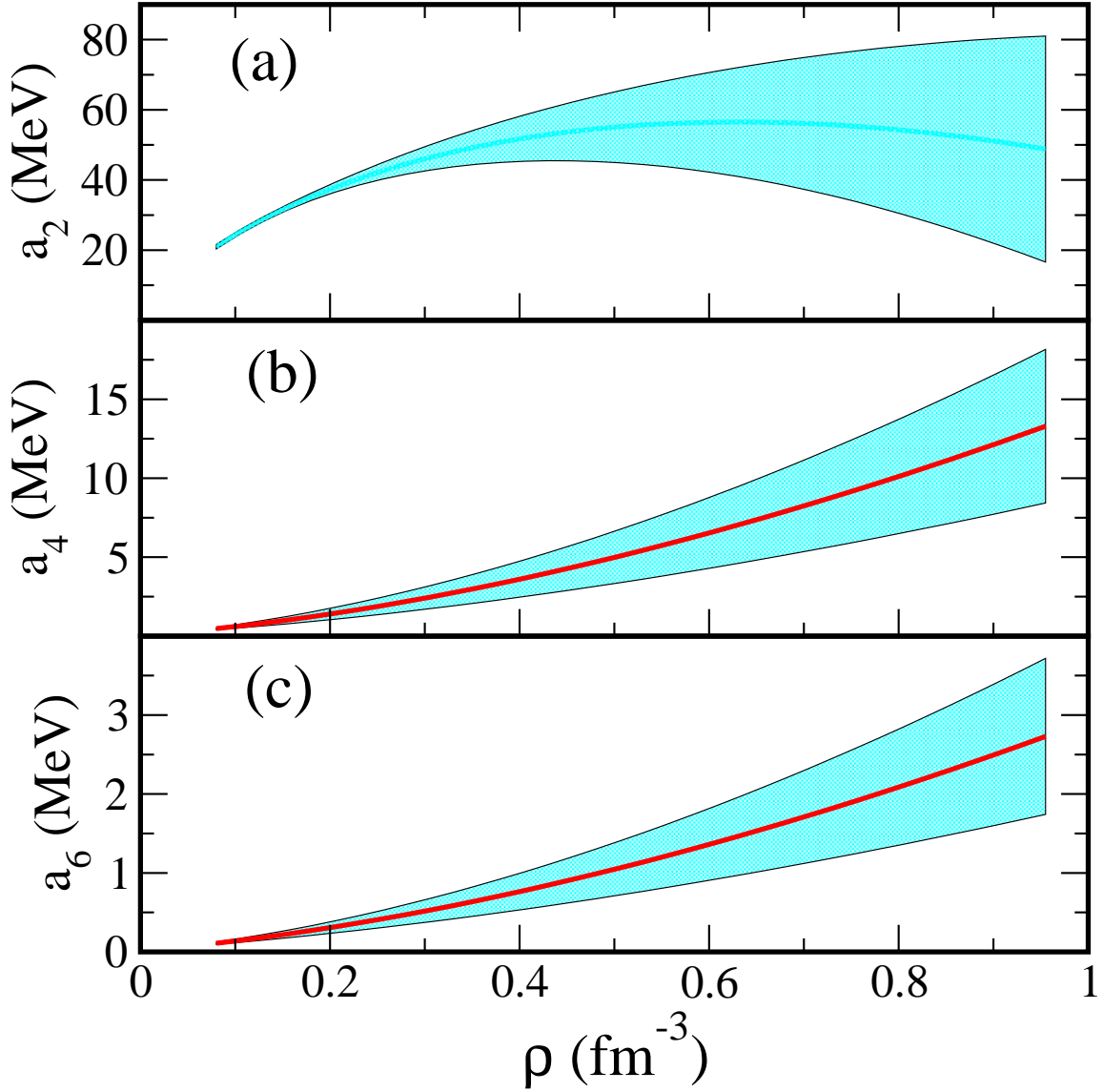}}
\caption{ (color online) The symmetry energy coefficients
$ a_2, a_4$ and $a_6$  displayed as a function of
density. Their central values are shown by the
red lines. The shaded regions are their uncertainties.}
\end{figure}

\subsection{The isovector elements of nuclear matter}

In Fig.~1, the symmetry coefficients $a_2, a_4, a_6$ as defined in
Eqs.(\ref{a22}),(\ref{a42}) and (\ref{a62}) are 
displayed as a function of density.
The coefficient $a_2$ increases with density  up to $\sim  4 \rho_0$, then
decreases slowly; $a_4$ and $a_6$, however, monotonically increase
with density. At the saturation density $\rho_0$ of symmetric nuclear matter, 
the values of $a_2, a_4$ and $a_6$ come out to be
$ 32.18\pm 0.78, 1.02\pm 0.23$ and $0.23\pm 0.04$ MeV, 
respectively. The higher order coefficients
are seen to be negligible at low densities, even around $\rho_0$
they are not appreciable validating the Bethe-Weiz\" acker conjecture.
The value of symmetry energy is seen to agree very well with the
estimate of 31$\pm $2 MeV extracted from a combination of various
experiments \cite{Tsang12,Lattimer13}.
At higher densities, the relative importance of the higher
order coefficients starts to show up. The shades in the figure refer
to the uncertainties in the coefficients which are quite significant 
as the density increases. The emergence of the relative importance of the
higher order coefficients with increasing density is shown in Fig.~2.
The growing difference of the total symmetry energy $e_{sym}$ (which
is the sum of all orders of the symmetry coefficients) from $a_2+a_4+a_6$
with density shows that still higher order terms need to be taken into 
consideration at very high densities and asymmetries  
prevalent near the core of the neutron star. The relatively smaller
values of the higher order symmetry coefficients in our calculation
at low densities and their growing importance with increasing density
are in fair agreement with those obtained from both non-relativistic
\cite{Constantinou14} and relativistic calculations \cite{Cai12}. Even
with reasonable variations of the empirical input data, no sizeable values 
for them are obtained near the normal density $\rho_0$.
At the highest density considered, the coefficient $a_4$ and $a_6$
are larger by about a factor of two in the present calculation as compared to 
those presented in reference \cite{Cai12} and \cite{Constantinou14} 
reminding us of the associated uncertainty in the calculated results 
in all models as one moves further away from the normal density around which 
the interaction parameters are determined.

The total density slope of symmetry energy $L_t$, the nuclear incompressibility
$K$ and its density derivative $M=3\rho dK/d\rho $ are displayed in
the three panels of Fig.~3 as a function of density. They grow with
density, so also their variances as shown by the shaded areas. 
The total symmetry density slope $L_t(\rho)$ is more relevant for asymmetric nuclear matter 
than the conventional $L(\rho)$. The pressure of neutron matter $P(\rho,\delta=1)$ is 
intimately related to $L_t(\rho)$ as $P(\rho,\delta=1)=P(\rho,\delta=0)+\frac{1}{3}\rho L_t(\rho)$. 
We have therefore chosen to display the density variation of $L_t(\rho)$ rather than 
that of $L(\rho)$ which is very similar.
At saturation density $L_{t,0}=63.8 \pm 8.6$ MeV, $L_0=58.5\pm 6.5$ MeV and  $K_0$ is the same
as the input value as it ought to be. The value of $L_0$ is seen to be somewhat lower than those 
obtained from earlier studies using different methodologies \cite{Agrawal12,Agrawal13,Vidana09,Colo14}, 
but is in good agreement with those obtained from fitting of selective experimental data on 
nuclear masses across the periodic table \cite{Mondal15, Mondal16} that includes highly neutron-rich 
nuclear systems. 
The value of incompressibility $K_c$ at a density $\rho_c (=0.71 \pm 0.005\rho_0 )$ is argued to
be more relevant \cite{Khan12,Khan13} as an indicator of the ISGMR
centroid. The incompressibility $K(\rho )$ calculated with a multitude
of EDFs of the Skyrme class, when plotted against density are seen to
cross close to this single density point $\rho_c$. The reported value
of $K_c \sim 35 \pm 4$ MeV \cite{Khan13} compares extremely well with
our calculated value of 34.1 $\pm $1.2 MeV. The computed value of $M_c$
$(=3\rho dK/d\rho |_{\rho_c})  =1062 \pm 102 $ MeV also compares very
favorably with $M_c=1050 \pm 100$ MeV \cite{Khan13} as obtained from the
analysis of known experimental ISGMR data. The value of $M_0 =M(\rho_0)$
at saturation density can not be compared with any benchmark value, but
since $M_0=12K_0+Q_0$ where $Q_0 =27 \rho_0^3\partial^3e(\rho ,0)/\partial
\rho^3 |_{\rho_0}$, $Q_0$ can be estimated (as $K_0$ is given). The
value $Q_0 =-360$ MeV conforms well with the one $Q_0=-350 \pm 30 $
MeV \cite{Chen09} obtained from examination of a host of standard Skyrme
interactions.  The evaluated value of symmetry incompressibility $K_\delta
=-382\pm 60$ MeV $(K_\delta =9\rho_0 ^2 \frac{\partial ^2e_{sym}}{\partial
\rho^2}|_{ \rho_0}- 6L_0 - Q_0L_0/K_0)$ is also in good consonance with
the reported value of $-370 \pm 120 $ MeV extracted from measurements
of isospin diffusion in heavy ion collisions  and with $\sim -350 $
MeV \cite{Pearson10} obtained from analysis of ISGMR data in Sn-isotopes.
The total uncertainties $\Delta X$ in the various observables $X$ are 
evaluated as \cite{Arfken05}, 
\bea
\Delta X= \sqrt{\sum_i (\Delta X_i)^2},
\eea
where $\Delta X_i=\frac{\partial X}{\partial Y_i}\Delta Y_i$, $\Delta X_i$ is 
the partial uncertainty induced by the uncertainty $\Delta Y_i$ in the input quantity 
$Y_i$ (say, 25 MeV in $K_0$). The derivatives $\frac{\partial X}{\partial Y_i}$ are 
calculated numerically.

It is worth mentioning at this juncture that the recent analyses
\cite{Jiang14,Jiang15,Wang15}    of the nuclear masses suggest a rather
high value for the fourth order coefficient (order with $\delta^4$)
of symmetry energy for finite nuclei. This coefficient is, however, not
to be equated with $a_4$ of Eq. (2), but possibly is indicative of the
term with $\delta^4$ ($E_{\rm sat,4})$, in the notation of \cite{Chen09}
in the series expansion in $\delta$ of the binding energy per nucleon
at saturation density of nuclear matter of asymmetry $\delta$.  It is
related to $a_4(\rho_0)$ as
 \begin{equation} E_{\rm sat,4} = a_4({\rho_0})
- \frac{L_0^2}{2K_0}.  \end{equation}
With values of $L_0 $ and $K_0$
in our model, this fourth order coefficient ($E_{\rm sat,4}$) is then
$\sim -6.7$ MeV. The magnitude of this coefficient may be compared with those obtained for a 
multitude of Skyrme interactions in Ref. \cite{Chen09} which is $\sim -4.6$ MeV.

\begin{figure}
\resizebox{6.0in}{!}{\includegraphics[]{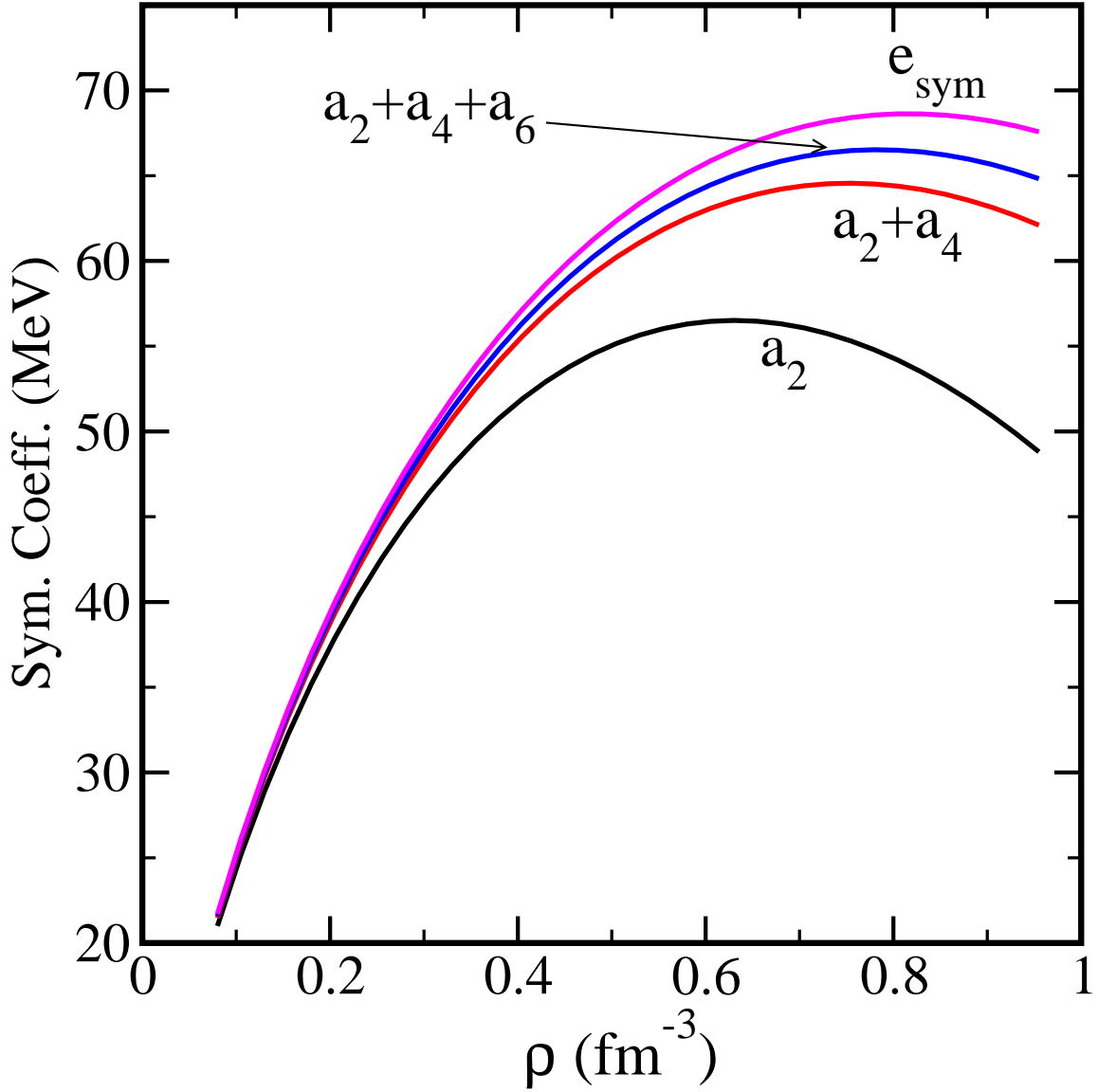}}
\caption{ (color online) The contributions of different orders of
symmetry energy coefficients $a_2, a_4, a_6$ to the total symmetry energy
coefficients $e_{sym}$ shown as a function of density.}
\end{figure}
\begin{figure}[b]
\resizebox{6.0in}{!}{\includegraphics[angle=-90]{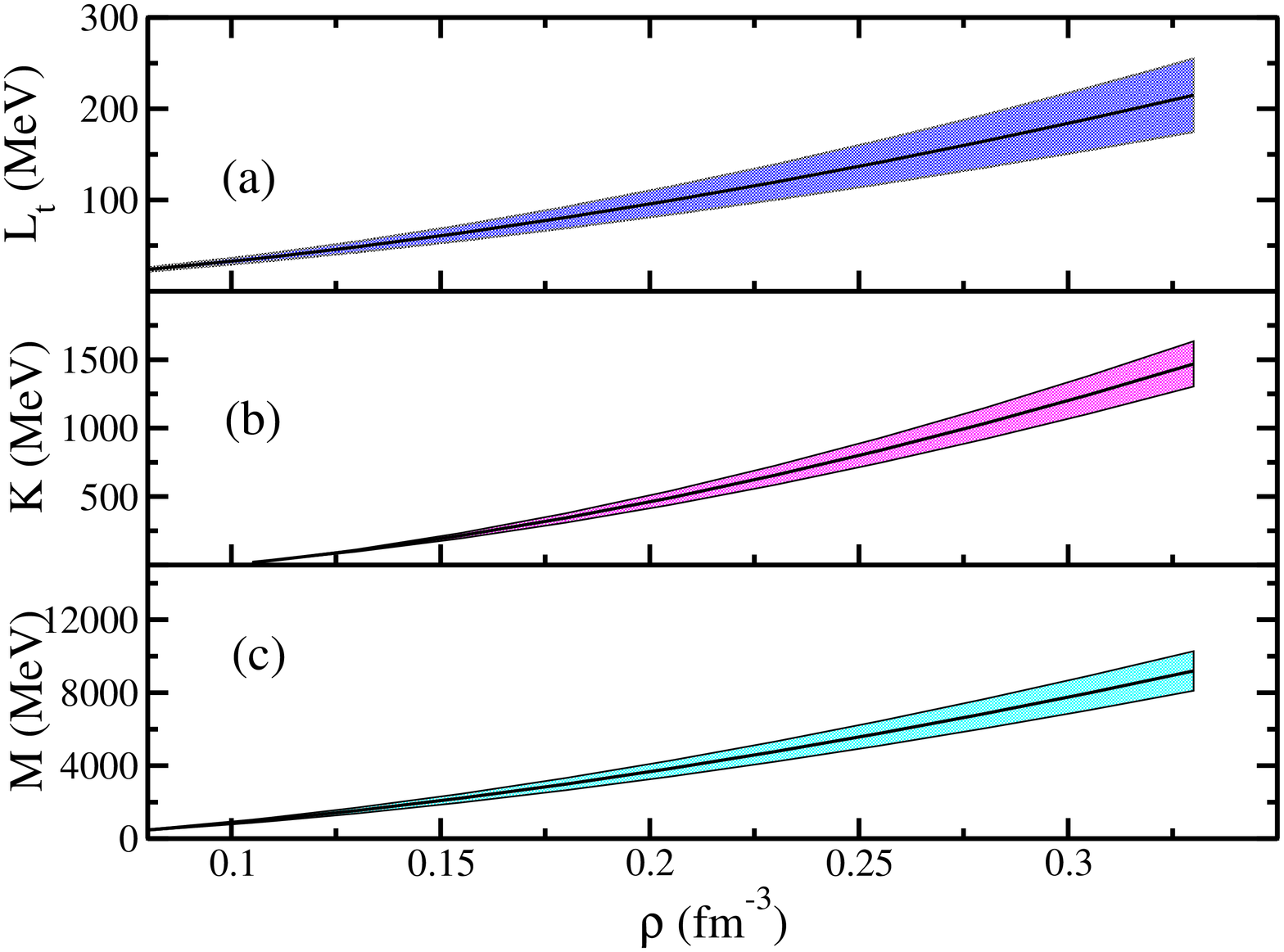}}
\caption{ (color online) The total density slope of symmetry energy $L_t$, the
incompressibility $K$ and its density derivative $M$ are shown in
panels (a), (b) and (c), respectively, at different values of density.
The full black line refer to their central values with the shaded regions
representing their uncertainties.}
\end{figure}

The isospin splitting of the nucleon effective mass is a useful
reference mark for an easy comprehension of the strength of the
momentum dependence of the nucleon isovector potential. This is 
still a poorly known quantity, 
even the signature of
the mass difference $(m_n^*-m_p^*)|_{\rho_0}$ is seen to be rather 
uncertain \cite{Ou11} within the Skyrme-Hartree-Fock approach.
There has been some recent interest in
understanding it from different perspectives. 
Analyzing comprehensive nucleon elastic scattering data over a 
wide energy domain for a large number of systems, Li $\it et~ al.$
\cite{Li15}
have reported a value for $(m_n^*-m_p^*)|_{\rho_0}/m$ =(0.41 $\pm $0.15)$\delta $
at saturation density. On the other hand, exploring the giant
resonances and the electric dipole polarizability in $^{208}$Pb,
a somewhat lesser value $(0.33\pm 0.16)\delta $
of the said isovector splitting is obtained \cite{Zhang15a}. From 
our calculation, it is easy to show, from Eq. (\ref {mstar}) and 
(\ref {vt1})  that at any density
\bea
\label{efmas1}
\frac{1}{m_n^*}-\frac{1}{m_p^*}=2V_n^1-2V_p^1      \nonumber \\
=\frac{8\pi}{b^2}(C_l-C_u)a^3\rho \delta
\eea
A little algebra leads to
\bea
\label{efmas2}
\frac{(m_n^*-m_p^*)}{m}=2\frac{K_2}{(K_1)^2}\left(\delta +\left(\frac{K_2}{K_1}\right)^2
\delta^3 + ....... \right),
\eea
where
\bea
\label{efmas3}
K_1=1+\frac{4\pi a^3}{b^2}m\rho (C_l+C_u),
\eea
and
\bea
\label{efmas4}
K_2=\frac{4\pi a^3}{b^2}m\rho (C_u-C_l).
\eea
At $\rho_0$, plugging in the values of the interaction parameters and
noting that the higher order terms in $\delta $ in Eq. (\ref {efmas2})
are negligible, we get,
\bea
\label{efmas5}
\frac{(m_n^*-m_p^*)}{m}|_{\rho_0} \simeq (0.209 \pm 0.017)\delta .
\eea

The results on the symmetry elements presented so far pertain to calculations 
with an energy density functional constructed with the momentum and density dependent 
SBM interaction, the parameters of which are fixed from empirical bulk nuclear data. 
To check the consistency of the results, the calculations have been repeated in the Skyrme framework. 
The energy per nucleon $e(\rho,\delta)$ in this framework is \cite{Brack85}
\bea
e(\rho,\delta)&=&g_1\left[\left(\frac{1+\delta}{2}\right)^{5/3}+\left(\frac{1-\delta}{2}\right)^{5/3}\right]\rho^{2/3} \nonumber\\
&&+\left(b_1+b_2\delta^2\right)\rho+\left(c_1+c_2\delta^2\right)\rho^{\gamma+1} \nonumber\\
&&+\left[d_1\left\{\left(\frac{1+\delta}{2}\right)^{5/3}+\left(\frac{1-\delta}{2}\right)^{5/3}\right\}\right.
\nonumber\\
&&\left.+d_2\left\{\left(\frac{1+\delta}{2}\right)^{8/3}+\left(\frac{1-\delta}{2}\right)^{8/3}\right\}\right]\rho^{5/3}.
\eea
The first term on the right-hand side is the free Fermi-gas energy, 
$g_1=\frac{\hbar^2}{2m}\frac{3}{5}(3\pi^2)^{2/3}=119.14$ MeV fm$^2$. There are seven 
parameters in this EDF, namely, $b_1, b_2, c_1, c_2, d_1, d_2$ and $\gamma$. The 
parameter $\gamma$ is related to the isoscalar bulk data $\frac{m_0^*}{m}, e_0, K_0$ and $\rho_0$ \cite{Agrawal05} 
as 
\bea
\gamma=\frac{-e_0-\frac{K_0}{9}+\left(\frac{4}{3}\frac{m}{m_0^*}-1\right)\frac{g_1\rho_0^{2/3}}{3\cdot 2^{2/3}}}
{e_0+\left(2\frac{m}{m_0^*}-3\right)\frac{g_1\rho_0^{2/3}}{3\cdot 2^{2/3}}}.
\eea
The isoscalar equations for $e_0, K_0$ and $P$(= 0 at $\rho_0$) yield the values of 
$b_1, c_1$ and $\left(d_1+\frac{d_2}{2}\right)$. For the remaining parameters, in addition to
the constraints $e_n$ for neutron matter at $\rho=0.1$ fm$^{-3}$, we need two other isovector
entities. We choose them to be $a_2(\rho_0)$ = 32.1 $\pm$ 0.31 MeV \cite{Jiang12} and
$a_2(\rho_1)$ = 24.1 $\pm$ 0.8 MeV \cite{Trippa08}. The former have been obtained recently from
a meticulous study of the double differences of "experimental" symmetry energies \cite{Jiang12},
the latter is obtained from giant dipole resonance analysis \cite{Trippa08}. All the other
empirical data are chosen to be the same as in the SBM framework. Equations for $a_2(\rho_0),\
a_2(\rho_1)$ and $e_n(\rho_1)$ yield the values of $b_2,\ c_2$ and $(d_1+2d_2)$. The values of
all the parameters entering the Skyrme EDF are thus known. Details about finding out the
parameters in the Skyrme framework are given in Ref. \cite{Alam14}. In Tab. \ref{tab2}, the
central values of the symmetry elements we deal with at and around the saturation density
obtained from the two frameworks are compared. They are compatible, 
the difference in the neutron-proton effective mass is seen to
be larger in the Skyrme prescription. Both are positive.
\begin{table*}[]
\caption{\label{tab2} Comparison of the values of the symmetry energy elements in the SBM and Skyrme framework 
at the densities indicated. }
\begin{ruledtabular}
\begin{tabular}{ccccccccccc}
 & $a_2(\rho_0)$ & $a_2(\rho_1)$ & $a_4(\rho_0)$ & $a_4(\rho_1)$ & $a_6(\rho_0)$ & $a_6(\rho_1)$ &
$L(\rho_0)$ & $L_t(\rho_0)$ & $K_{\delta}$ & $\left(\frac{m^*_n-m^*_p}{m}\right)_{\rho_0}$ \\
\hline
SBM & 32.2 & 24.4 & 1.01 & 0.61 & 0.23 & 0.14 & 58.3 & 63.8 & -382 & 0.21$\delta$ \\
\hline
Skyrme & 32.1 & 24.1 & 1.47 & 0.83 & 0.25 & 0.14 & 57.3 & 65.6 & -412 & 0.49$\delta$ \\
\end{tabular}
\end{ruledtabular}
\end{table*}

\subsection{Supranormal density and neutron stars}
Now that the EDF so constructed in the SBM framework produces results that are in reasonably good
agreement with those obtained from different perspectives (both in
experiment and theory) at normal and subnormal densities, it would
be interesting to see how the EDF behaves at high densities, how
the pressure changes as a function of  baryon density and asymmetry. The
baryon pressure is an essential element in shaping properties of
neutron star matter, in understanding the lower limit of
the maximum mass of neutron star $M_{max}^{NS}$. 
Our calculations show that at low densities, 
the pressure for SNM is lower compared to that for PNM, however, it catches
up at higher densities. This crossing density is found to be dependent
on the value of $m_0^*/m$. For higher values of effective 
mass, the crossing density
is lower ($\sim 3\rho_0 $ for $m_0^*/m =0.75$), but moves up as $m_0^*/m$
decreases ($\sim 5.5 \rho_0 $ for $m_0^*/m=0.65 $). In Fig.~4, the pressure-
density relation is portrayed in the upper panel for SNM and in the lower
panel for PNM with $m_0^*/m$=0.65$\pm 0.1$. The violet shades show
the calculated uncertainties. The shaded red and orange regions in the
upper  panel display the 'experimental' EoS for SNM extracted from 
collective flow data \cite{Danielewicz02} and from data for Kaon 
production \cite{Fuchs06,Fantina14}, respectively. The shaded green region 
in the EoS of PNM is a theoretically obtained result where the density
dependence of symmetry energy is taken to be soft. The red shaded
region is the one where the said density dependence is modeled as stiff
\cite{Prakash88}. These regions have  very good overlap with the one 
obtained from our calculation.

\begin{figure}
\resizebox{5.0in}{!}{\includegraphics[]{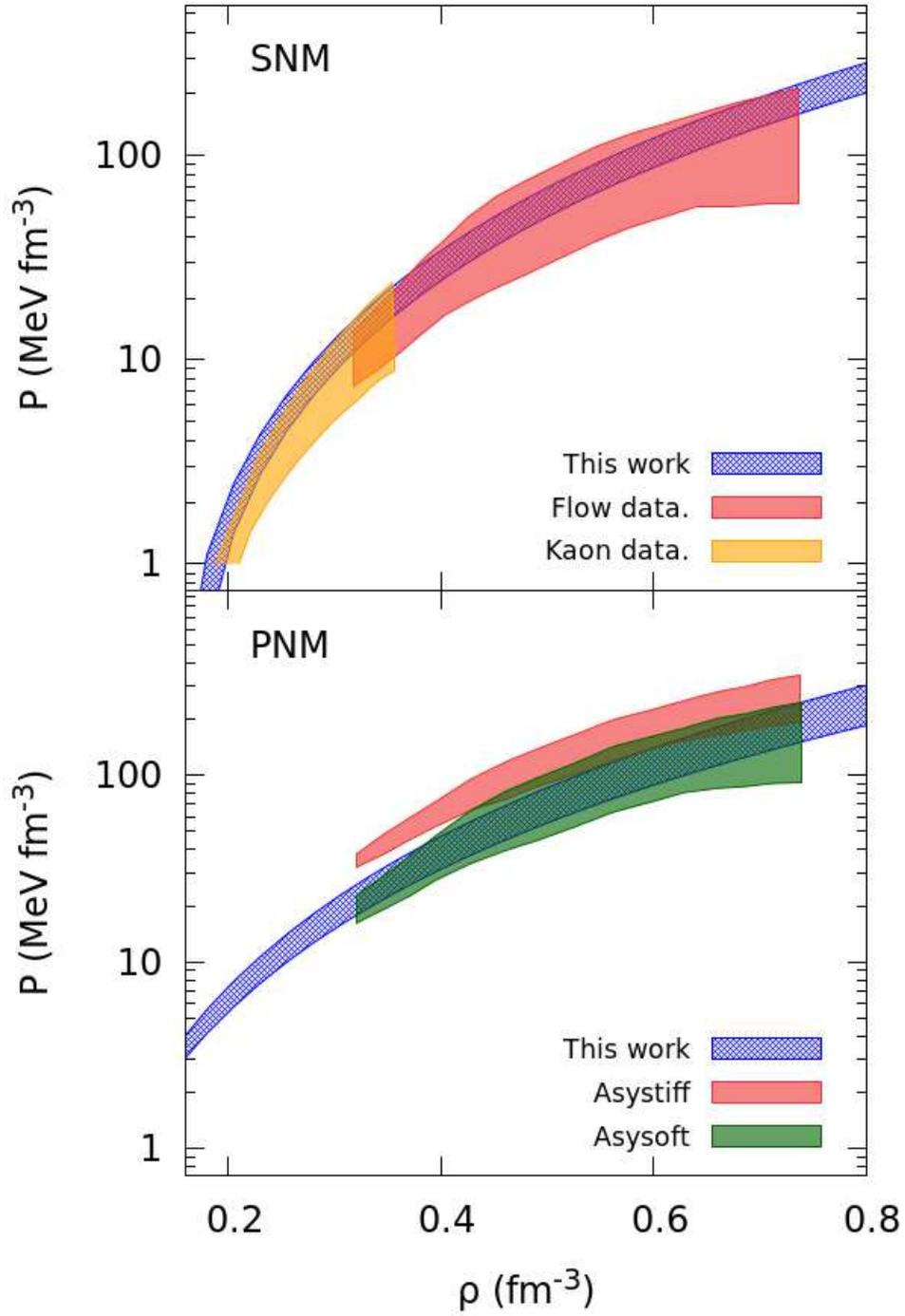}}
\caption{ (color online) The EoS for symmetric nuclear matter
(upper panel) and for pure neutron matter (lower panel). The calculated
results and the experimental data are as indicated. See text for details .}
\end{figure}

Solving the general relativistic Tolman-Oppenheimer-Volkoff (TOV) equation
\cite{Weinberg72}, we have calculated $M_{max}^{NS}$ for neutron star
with different values of $m_0^*/m$. The EoS for the crust was taken from
the Baym, Pethick and Sutherland model \cite{Baym71}. The EoS for
the core region was calculated under the assumption of a charge neutral 
uniform plasma of neutrons, protons, muons and electrons in
$\beta-$ equilibrium. 
Possible phase transition to exotic phases such as hyperons, kaons etc.
at high densities softens the EoS somewhat. This is not taken into 
consideration here.
The maximum mass  calculated within this framework is shown in 
Fig.~5 as a function of $m_0^*/m$. We note that $M_{max}^{NS}$ goes up
with decreasing $m_0^*/m$. At $m_0^*/m=0.65$, the calculated value
for $M_{max}^{NS} =(1.95 \pm 0.14)M_{\odot}$ is consistent with the 
currently observed values of $(1.97\pm 0.04)M_{\odot}$ for the 
pulsar PSR J1614-2230 \cite{Demorest10} and also with  the value of 
$(2.01\pm 0.04)M_{\odot}$ \cite{Antoniadis13} .

\begin{figure}
\resizebox{6.0in}{!}{\includegraphics[]{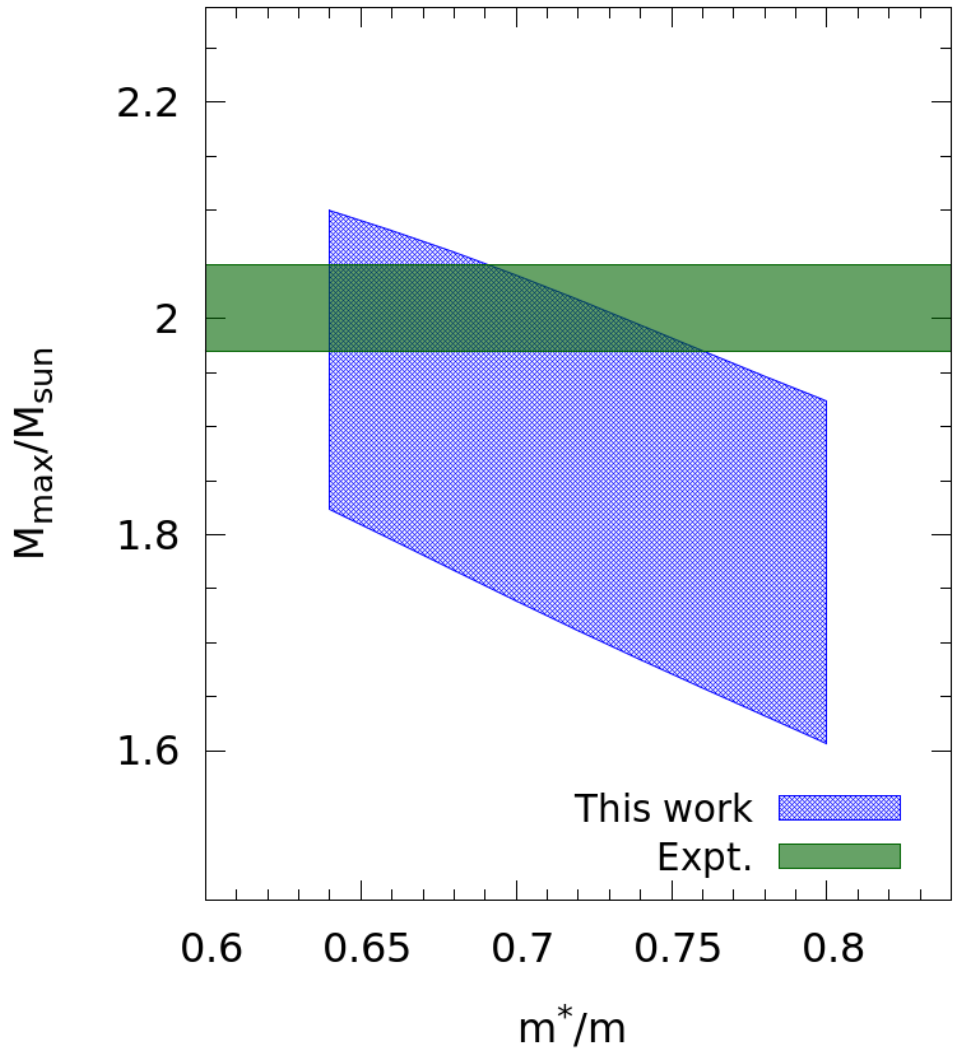}}
\caption{(color online) Dependence of the lower limit of the maximum
mass of neutron star $M_{max}^{NS}$ (in units of solar mass $M_{\odot }$)
on the effective nucleon mass for symmetric nuclear matter.}
\end{figure}

\section{Concluding remarks}

From well-constrained empirical data relevant for nuclear matter at
saturation and subsaturation densities, we have constructed an energy
density functional based on a finite-range, momentum and density dependent
interaction. The different elements related to symmetry energy and
their density dependence are then analyzed with this EDF. The density
slope of symmetry energy $L(\rho )$, the density dependence of nuclear
incompressibility $K(\rho )$, its density slope $M(\rho )$, the symmetry
incompressibility $K_{\delta}$ at saturation for asymmetric matter $-$ all
these are seen to be in excellent agreement with their recently obtained
values from different perspectives. Calculations done in a Skyrme-inspired
framework for the EDF with the same input empirical data do not change
the conclusions much.  We modeled the EoS with the SBM EDF so that the
calculated $M_{max}^{NS}$ conforms well with the experimentally observed
one; keeping this in mind, still it must be said that the agreement of
our constructed EoS with the 'experimental' one over an extended density
plane is very striking. From this overall consistency of our constructed
EoS and the derivative results built from empirical data, we infer that
the higher order symmetry coefficients $a_4, a_6$ etc. of infinite nuclear
matter are not sizeable at and around saturation density, but grow with
increasing density.  This is in fair agreement with earlier investigations
\cite{Constantinou14,Cai12,Seif14} and confirms that the EoS of asymmetric
nuclear matter, though conforms to the parabolic approximation at normal
and sub-saturation density deviates significantly from it as the density rises.

In calculating the maximum mass $M_{max}^{NS}$ for neutron star, we have
confined ourselves to homogeneous nuclear matter in $\beta-$equilibrium.
Exotic degrees of freedom near the interior of the star may change the
calculated value of $M_{max}^{NS}$ somewhat, this has been left out of
our consideration in the present description.

\section{Acknowledgments}

J.N.D. acknowledges support from the Department of Science and Technology,
Government of India. The assistance of Tanuja Agrawal in the preparation
of the manuscript is gratefully acknowledged.

\appendix
\section{}
Here we show how the parameters of the interaction are determined.  The
single-particle potentials $V_0, V_1, V_2$ 
and the effective mass $m_0^*$ refer to the entities for SNM
at the saturation density $\rho_0$. 

From Eq. (\ref{etot2}), we know $V_0$ from the empirical inputs,
\begin{equation}
V_0 = 2e_0 - \frac{3}{5}\left\{ \frac{P_{F,0}^2}{2m}(1 + \frac{m}{m_0^*})
\right \}, 
\end{equation}
where $P_{F,0}$ is the Fermi momentum obtained from 
\begin{equation}
\rho_0 = \frac{2P_{F,0}^3}{3\pi^2\hbar^3}.
\end{equation}

The momentum dependent part $V_1$ is known from 
\begin{equation}
V_1=\frac{1}{2m} \left (\frac{m}{m_0^*} - 1\right ),
\end{equation}
which for symmetric matter is [Eq. (\ref{vt1})]
\begin{equation}
V_1=\frac{2\pi a^3}{b^2}(C_l + C_u)\rho_0.
\end{equation}
The rearrangement term $V_2$ and the potential $V_0$ for symmetric matter
are 
\begin{eqnarray}
V_2 = \pi a^3 d^2 (2\rho_0)^\alpha \alpha (C_l+C_u)\rho_0,\\
V_0 = V_1\left [\frac{3}{5}P_{F,0}^2 - b^2\left \{ 1 - d^2
(2\rho_0)^\alpha\right \} \right ].
\end{eqnarray}
Eqs. (A4 - A6) give a relation between the three potentials, 
\begin{equation}
V_0 = V_1\left [\frac{3}{5}P_{F,0}^2 - b^2\right ]+ \frac{2V_2}{\alpha}.
\end{equation}
From Eq. (\ref{pres2}), $V_2$ at saturation (where pressure is zero) 
can also be calculated in terms
of known quantities 
\begin{equation}
V_2 =-e_0 + \frac{P_{F,0}^2}{10m} (3- 2 \frac{m}{m_0^*}).
\end{equation}

From Eqs. (A4) and (A5) 
\begin{equation}
d^2 (2\rho_0)^\alpha = \frac{2V_2}{\alpha b^2 V_1}.
\end{equation}
Putting this in Eq. (\ref{comp2}) gives 
\begin{equation}
K_0 = -3V_0 + V_2 (9\alpha + \frac{9}{\alpha} + 12) + V_1(10.8 P_{F,0}^2 -
4.5 b^2).
\end{equation}
Eqs. (A7) and (A10) give $b$ and $\alpha$. Eq. (A9)  then gives  $d^2$.
The values of $\alpha, b^2$ and $d^2$ are given as,
\begin{eqnarray}
\alpha&=&\frac{1}{V_2}\left [ ( \frac{K_0}{9} - \frac{V_0}{6} -
\frac{9}{10}P_{F,0}^2 V_1)\right ] - \frac{4}{3},\\
b^2&=&\frac{3}{5}P_{F,0}^2 + \frac{1}{V_1}\left [ \frac{2V_2}{\alpha} -
V_0 \right ], \\
d^2&=&\frac{2V_2}{\alpha b^2(2\rho_0)^\alpha V_1}.
\end{eqnarray}

The value of  $C_la^3$ is determined from Eq. (\ref{en1}). 
Eqs. (A3) and (A4)  then gives $C_u a^3$. 
\section{}
In the framework of the effective interaction chosen, the effective mass
$m_0^*/m$ is seen to have a lower bound. In Eq. (A10), putting the value of
$b^2 V_1$ from Eq. (A7) we have, 
\begin{equation}
\alpha = \frac{[K_0 - 1.5 V_0 - 12 V_2 - 8.1 P_{F,0}^2V_1]}{9V_2}.
\end{equation}
With values of $V_0, V_1$ and $V_2$ from Eqs. (A1), (A3) and (A8), one
gets an equation for  $\alpha$ from Eq. (B1)  in terms of empirical
quantities, 
\begin{equation}
\alpha = \frac{\left [K_0 - 3e_0 + \frac{9P_{F.0}^2}{20m}(1 +
\frac{m}{m_0^*}) - 8.1 \frac{P_{F,0}^2}{2m}(\frac{m}{m_0^*} - 1)\right
]}{9V_2}-\frac{4}{3}.
\end{equation}

With given values of $K_0, e_0, \rho_0$ etc., examination of Eq. (B2) shows
that as $m_0^*/m$ starts decreasing from unity, the value of $\alpha$
starting from a positive value 
become lower and lower until at some value of $m_0^*/m$, it crosses zero
and then becomes negative. The value of $b^2$ then makes a sudden
transition from a large positive value to a large negative value. Since
the density-dependent part  of the interaction 
\begin{equation}
d^2 (2\rho_0)^\alpha = \frac{2V_2}{\alpha b^2 V_1} 
\end{equation}
should be repulsive and should increase with density, $\alpha$ should be
positive; the physically accepted 
minimum value  of $m_0^*/m$ is then determined from the
condition ($V_2 $ is still finite from the empirical inputs)
\begin{equation}
K_0 - 1.5V_0 - 12 V_2 - 8.1 P_{F,0}^2V_1 = 0, 
\end{equation}
which yields 
\begin{equation}
\left ( \frac{m_0^*}{m}\right )_{\rm min} =\frac{
\frac{6P_{F,0}^2}{m}}{45e_0+5K_0 +4.5 \frac {P_{F,0}^2}{m}}.
\end{equation}
With the values of the empirical quantities chosen,
$\left ( \frac{m_0^*}{m}\right )_{\rm min}$ is $\sim 0.64$.


\end{document}